\documentclass[aps,prl,shownopacs,superscriptaddress,twocolumn]{revtex4}

\usepackage{siunitx}
\usepackage{bm}
\usepackage{ifpdf}
\usepackage{float}
\usepackage{upgreek}
\usepackage{xcolor}
\usepackage{graphicx}
\usepackage{hyperref}
\usepackage{amsmath}
\usepackage{epstopdf}
\usepackage{bbm}
\usepackage{braket}

\usepackage{natbib}

\makeatletter
\def\NAT@def@citea{\def\@citea{\NAT@separator}}
\makeatother

\usepackage{tabularx}
\usepackage[normalem]{ulem}
\usepackage[T1]{fontenc}
\usepackage{gensymb}
\usepackage{amsmath,amssymb,amsfonts}
\usepackage{amsthm}
\newcommand{\Tr}{\ensuremath{\,\mathrm{Tr}}}
\newcommand{\ketbra}[2]{|#1\rangle\!\langle#2|}
\usepackage[bb=boondox]{mathalfa}
\DeclareFontFamily{U}{futm}{}
\DeclareFontShape{U}{futm}{m}{n}{
  <-> s * [.95] fourier-bb
  }{}
\DeclareSymbolFont{Ufutm}{U}{futm}{m}{n}
\DeclareSymbolFontAlphabet{\mathbb}{Ufutm}
\DeclareMathAlphabet{\mathbbb}{U}{bbold}{m}{n}

\begin{document}

\title{Quantum cryptography integrating an optical quantum memory}

\author{Hadriel Mamann}
\affiliation{Laboratoire Kastler Brossel, Sorbonne Universit\'{e}, CNRS, ENS-Universit\'{e} PSL, Coll\`{e}ge de France, 4 Place
Jussieu, 75005 Paris, France}
\author{Thomas Nieddu \footnotemark[2]\footnotetext{\footnotemark[2]Present address: Welinq, Paris, France.}}
\affiliation{Laboratoire Kastler Brossel, Sorbonne Universit\'{e}, CNRS, ENS-Universit\'{e} PSL, Coll\`{e}ge de France, 4 Place
Jussieu, 75005 Paris, France}
\author{F\'{e}lix Hoffet \footnotemark[3]\footnotetext{\footnotemark[3]Present address: ICFO - Institut de Ciencies Fotoniques, The Barcelona Institute of Science and Technology, Barcelona, Spain.}}
\affiliation{Laboratoire Kastler Brossel, Sorbonne Universit\'{e}, CNRS, ENS-Universit\'{e} PSL, Coll\`{e}ge de France, 4 Place
Jussieu, 75005 Paris, France}
\author{Mathieu Bozzio}
\affiliation{University of Vienna, Faculty of Physics, Vienna Center for Quantum Science and Technology (VCQ), 1090 Vienna, Austria}\author{F\'{e}lix Garreau de Loubresse}
\affiliation{Laboratoire Kastler Brossel, Sorbonne Universit\'{e}, CNRS, ENS-Universit\'{e} PSL, Coll\`{e}ge de France, 4 Place
Jussieu, 75005 Paris, France}
\author{Iordanis Kerenidis}
\affiliation{Universit\'e de Paris, CNRS, IRIF, 75013 Paris, France}
\author{Eleni Diamanti}
\affiliation{LIP6, CNRS, Sorbonne Universit\'{e}, 75005 Paris, France}
\author{Alban Urvoy}
\affiliation{Laboratoire Kastler Brossel, Sorbonne Universit\'{e}, CNRS, ENS-Universit\'{e} PSL, Coll\`{e}ge de France, 4 Place
Jussieu, 75005 Paris, France}
\author{Julien Laurat}
\email{julien.laurat@sorbonne-universite.fr}
\affiliation{Laboratoire Kastler Brossel, Sorbonne Universit\'{e}, CNRS, ENS-Universit\'{e} PSL, Coll\`{e}ge de France, 4 Place
Jussieu, 75005 Paris, France}

\date{\today}

\maketitle

\textbf{Developments in scalable quantum networks rely critically on optical quantum memories, which are key components enabling the storage of quantum information. These memories play a pivotal role for entanglement distribution and long-distance quantum communication, with remarkable advances achieved in this context. However, optical memories have broader applications, and their storage and buffering capabilities can benefit a wide range of future quantum technologies. Here we present the first demonstration of a cryptography protocol incorporating an intermediate quantum memory layer. Specifically, we implement Wiesner's unforgeable quantum money primitive with a storage step, rather than as an on-the-fly procedure. This protocol imposes stringent requirements on storage efficiency and noise level to reach a secure regime. We demonstrate the implementation with polarization encoding of weak coherent states of light and a high-efficiency cold-atom-based quantum memory, and validate the full scheme. Our results showcase a major capability, opening new avenues for quantum memory utilization and network functionalities.\\}

Considerable efforts have been dedicated to the development of optical quantum memories using a variety of physical platforms, ranging from single emitters to atomic ensembles \cite{Heshami2016,Lei2023}. These advancements are driven by the promise of distributing quantum resources between remote locations using quantum repeater architectures, ultimately building a future quantum internet \cite{Kimble2008,Sangouard2011,Wehner2018,Azuma2023}. Remarkable demonstrations, both in laboratory settings and deployed telecom fiber networks, are paving the way for this ambitious goal \cite{Chou2007,Yu2020,Lago2021,Krutyanskiy2023, Liu2024,Knaut2024, Stolk2024}.

While entanglement distribution has been a primary focus in optical quantum memory research \cite{Lei2023}, broader applications for quantum technology have remained largely unexplored. Among the various storage platforms, some of them are absorptive quantum memories, capable of storing an incoming optical quantum state and retrieving it on demand \cite{Lei2023}. They are thereby critical devices that can be used for resource synchronization and general networking operations. Examples of future use cases include buffering quantum data alongside quantum processor units or along a transmission line. These fundamental operations impose stringent and challenging constraints on memory performance, particularly in terms of efficiency and noise minimization \cite{Singh2023}.

\begin{figure}[b!]
\centering
\includegraphics[width=0.9\columnwidth]{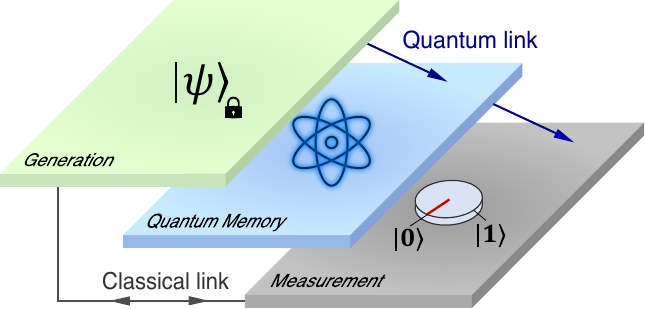}
\caption{\textbf{Quantum cryptographic protocol with an intermediate quantum memory layer.} In future quantum networks and use cases as unforgeable quantum money, optical memories, which allow data to be stored and retrieved on demand, play a central role. The incorporation of these memories puts stringent constraints on secure operation regions in terms of storage-and-retrieval efficiency and added noise.}
\label{figure1}
\end{figure} 

Here we present the first realization of a cryptographic primitive that incorporates an optical quantum memory layer, as illustrated in Fig. \ref{figure1}. Specifically, we implement the unforgeable quantum money protocol, a foundational scheme in the quantum cryptography field, originally proposed by Wiesner \cite{Wiesner83,Diamanti2016}. In this protocol, a central authority issues banknotes, credit cards or tokens comprised of quantum states, whose unforgeability is intrinsically guaranteed by the no-cloning theorem. The protocol is designed to prevent malicious clients and intermediate parties from double-spending the originally entitled value. 

To date, optical implementations have demonstrated the on-the-fly generation and verification of quantum money \cite{Bozzio2018,Guan2018}, omitting the crucial intermediate quantum storage step, which allows for spending flexibility. While interesting alternatives have been proposed, such as replacing quantum storage with a network of trusted agents \cite{Kent2020,Jiang2024} or use-cases where flexibility is not required \cite{Schiansky2023}, general applications call for on-demand storage and retrieval of quantum money. In our experiment, we demonstrate this combination and rigorously characterize the security threshold of the complete operation. This achievement was made possible by the use of a quantum memory based on an ensemble of laser-cooled neutral atoms, leveraging the high performance metrics -- close-to-unity efficiency and very low noise -- recently obtained with this platform \cite{VernazGris2018,Wang2019,Cao2020}.

\begin{figure}[t!]
\centering
\includegraphics[width=0.9\columnwidth]{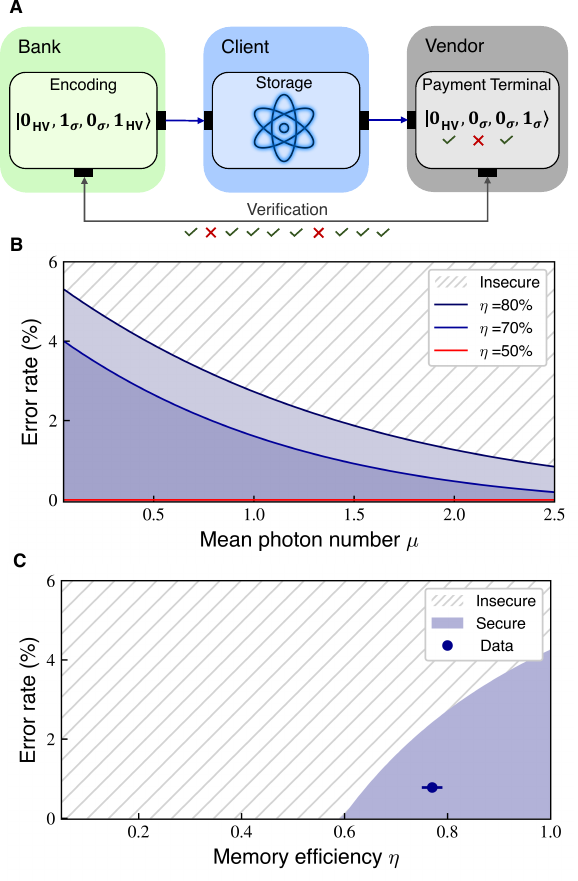}
\caption{\textbf{Quantum money protocol with memory storage and retrieval.} (\textbf{A}) The bank encodes a random secret key into a sequence of polarisation qubits chosen from two bases, $\{\ket{H},\ket{V}\}$ or $\{\ket{\sigma^+},\ket{\sigma^-}\}$, and stores them into a quantum memory provided to the client. In a transaction, the client retrieves the states from the memory and hands them to the vendor who performs the measurement in one of the encoding bases. For verification, the vendor communicates the measurement results and the chosen basis to the bank, allowing to calculate the error rate $\varepsilon$. (\textbf{B}) The communication is considered secure if the error rate falls below a specified threshold (solid lines), which is highly dependent on the mean photon-number per pulse $\mu$ for weak coherent states and on the memory efficiency $\eta$. (\textbf{C}) For a typical mean photon-number per pulse $\mu=1$, a successful protocol (shaded area) requires high efficiency and low error rate. The blue point indicates our experimental result.}
\label{figure2}
\end{figure}

The practical implementation is illustrated in Fig. \ref{figure2}a. This scheme consists of four steps. First, the bank encodes a uniformly random secret key onto a sequence of quantum bits (qubits), using conjugate codings \cite{Wiesner83,BB84}. In our work, the encoding is realized in polarization and the bases are either linear $\bigl\{\ket{H},\ket{V}\bigr\}$ or circular polarizations $\bigl\{\ket{\sigma^+},\ket{\sigma^-}\bigr\}$. Second, the qubits are stored into a quantum memory, materializing here the quantum credit card held by the client. In a third spending step, the client retrieves the data from the memory and forwards them to a vendor, who measures each qubit in one of the two polarization bases, chosen randomly. In the final verification step, the vendor classically communicates the basis choice and the associated measurement results to the bank, which checks for consistency with the original key. This single-round process provides the error rate $\varepsilon$ of the transaction. In this scenario, both the bank and the vendor are trusted.

In an ideal case, measuring a non-zero error rate would immediately signal an unauthorized double-spending attempt to the honest parties. In the presence of noise and finite channel efficiencies, some fraction of experimental imperfections should be tolerated for a practical protocol to succeed. However, a malicious party may in turn exploit this fault tolerance to hide their double-spending attempts. Moreover, the implementation relies on weak coherent states, which enables additional attacks such as photon-number splitting and unambiguous state discrimination due to the Poisson photon statistics \cite{Scarani2009}. This calls for a rigorous security analysis, identifying a combination of noise, losses and mean photon number for which no malicious party is able to successfully cheat \cite{BozzioPRA}. We detail such an analysis in the Supplementary Information, identifying the optimal quantum cloning strategy that minimizes the noise and losses introduced by the adversary \cite{Molina2013}. 

The resulting error-rate thresholds are given in Fig. \ref{figure2}b as a function of the mean photon number for different values of memory storage-and-retrieval efficiency. The areas above the thresholds are insecure. We first observe that, due to the no-cloning theorem, an efficiency above 50\% is required to have a possible range of secure operations. Then, as efficiency increases, the threshold rises. However, this is counterbalanced by a decrease for higher mean photon numbers. For a typical mean photon number $\mu=1$, Fig. \ref{figure2}c provides the secure operation regime as a function of error rate and efficiency. As depicted, this regime occupies a small corner of the parameter space and is challenging to achieve. It imposes stringent requirements on the memory layer, made possible only by recent advancements in the field, as demonstrated here for the first time.

\begin{figure*}[t!]
\vspace{-0.3cm}
\includegraphics[width=1.9\columnwidth]{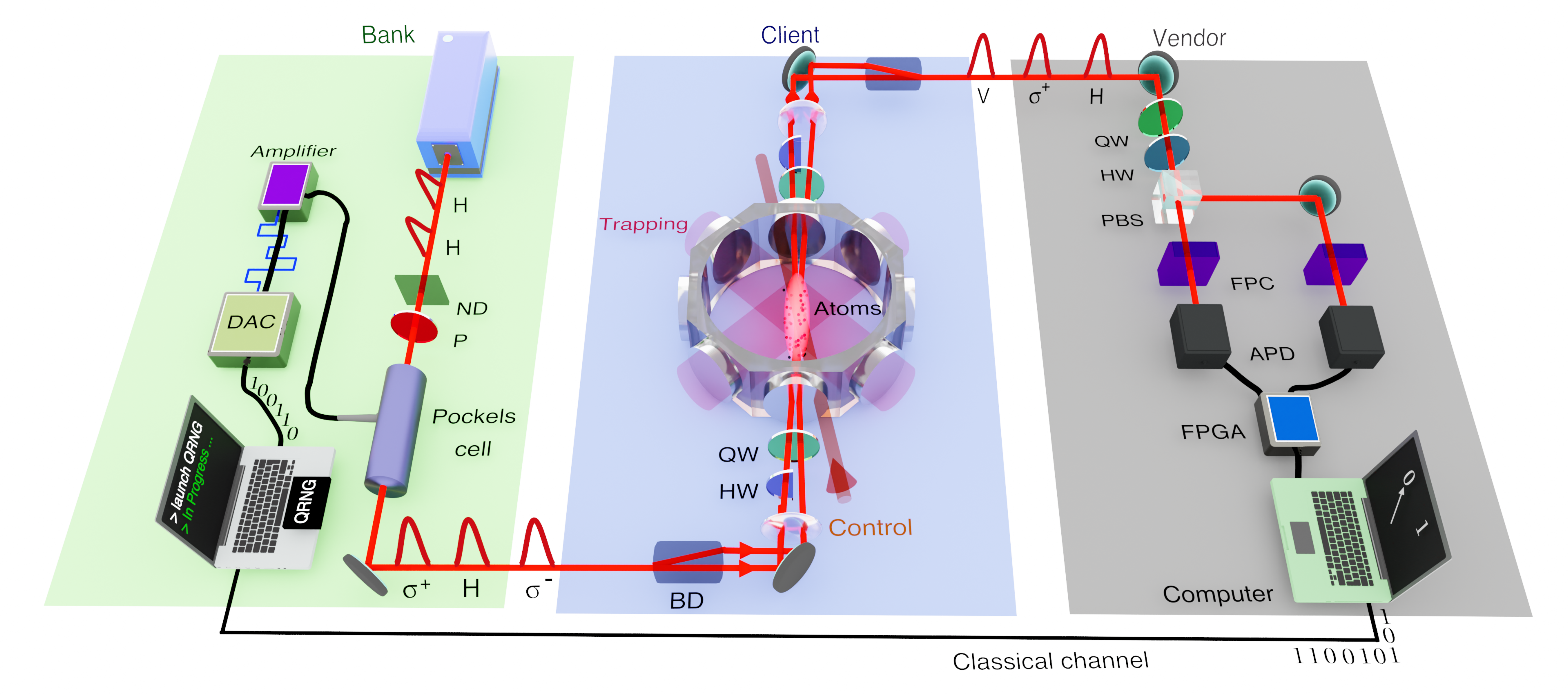}
\vspace{-0.2cm}
\caption{ \textbf{Experimental setup.} The three panels illustrate the encoding process (bank), the transmission line incorporating a quantum memory (client), and the detection stage (vendor). A quantum random number generator (QRNG) generates a secret key, which is used after voltage conversion (DAC) and amplification to prepare polarization states encoded on weak coherent states via a Pockels cell. The qubit states are then stored in a quantum memory based on an elongated ensemble of laser-cooled cesium atoms with ultra-high optical depth. An additional laser field dynamically controls the reversible mapping. To optimize storage, the polarization qubits are first converted into dual-rail qubits using a beam displacer (BD), and the reverse process is performed after retrieval. At the final stage, the polarisation states are measured in a chosen $\bigl\{H,V\bigr\}$ or $\bigl\{\sigma^+,\sigma^-\bigr\}$ basis using waveplates (QW, HW), a polarizing beam splitter (PBS), and two single-photon avalanche photodiodes (APD). Fabry-Perot cavities (FPC) are employed to filter the residual control beam leakage. The error rate $\varepsilon$ is determined by comparing the acquired data to the secret key through a classical channel.}
\label{figure3}
\end{figure*}

The experimental setup is detailed in Fig. \ref{figure3}. To generate the optical qubits, we prepare weak coherent-state pulses at the single-photon level and encode their polarization using a Pockels cell, with the encoding choice driven by a quantum random number generator. Four distinct bit combinations determine the basis and state, which are then converted to voltages and amplified. To meet the stringent requirement on the error rate, we notably achieved over 99.5\% polarization fidelity, independent of the encoded state thanks to a specific temporal sequence optimization of the Pockels cell (see Supplementary Information), outperforming the previous on-the-fly implementation \cite{Bozzio2018}. 

The optical qubits are then stored in a quantum memory based on a large ensemble of laser-cooled cesium atoms (see Supplementary Information). Polarization storage is achieved using a dual-rail configuration, which involves two paths within the ensemble, one corresponding to the H polarization and the other to the V polarization. This configuration is implemented via two beam displacers, one placed before the memory and the other after for path recombination, forming a passively stable interferometer. This stability arises from the limited dimension of the system and the inversion of the short and long paths in the displacer media \cite{VernazGris2018,Matsukevich2004,Chou2007,Laurat2007}. 

A critical parameter for achieving high-efficiency storage is the optical depth (OD) of the atomic ensemble. We implemented a compressed two-dimensional magneto-optical trap that enables to reach an OD up to 400 (see Supplementary Information) \cite{VernazGris2018}. The cesium atoms are initially pumped into the ground state $\big|g\bigr\rangle=\big|6S_{1/2},F=3\bigr \rangle$. The optical pulses are stored using the dynamic electromagnetically induced transparency (EIT) scheme with an additional laser beam called the control, phase-locked with the signal (see Supplementary Information). The control beam is turned on before the arrival of the signal and turned off when the pulse is fully compressed into the cloud, thereby converting coherently the optical qubit into a long-lived atomic collective excitation. All the experiment is performed on the cesium $D_1$ line, which is essential for achieving high efficiency as it limits off-resonant excitations and decoherence during the mapping process~\cite{Cao2020} . 

After a defined storage time, to read out the memory, the client turns on the control beam and hands over the qubits to the vendor who measures them in one of the two specific polarization bases, via a half waveplate, a quarter waveplate and a polarizing beam splitter. At this stage, the security of the protocol can be verified. 

\begin{figure}[t!]
\includegraphics[width=0.98\columnwidth]{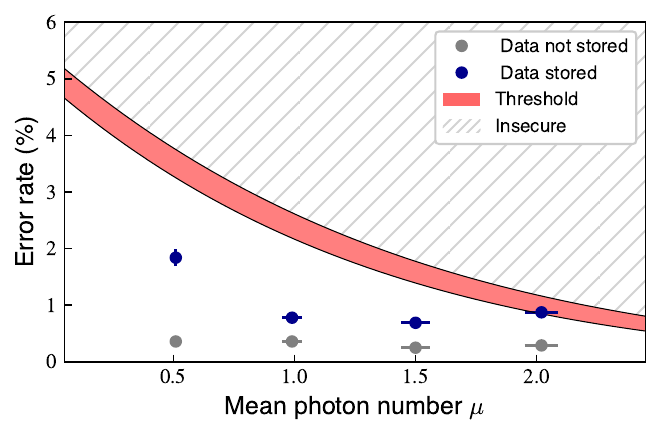}
\caption{\textbf{Experimental results and security threshold.} The error rates are shown for different mean photon numbers per pulse $\mu$, without  storage indicated in grey and with intermediate storage in blue. These rates are calculated as the average of error rates for the two measurement bases. The light red area represents the security threshold determined for a measured average efficiency of $\eta=(77\pm2)\%$ across the four mean photon numbers. Error bars for the error rates account for the statistical uncertainty of photon counts while error bars on the mean photon numbers correspond to power fluctuations during the overall data acquisition.}
\label{figure4}
\end{figure}

We now turn to data analysis, which consists in comparing the data collected by the vendor with the secret key generated by the bank. Verification is only performed on qubits initially encoded in the measurement basis chosen by the vendor, with the others excluded. This is akin to the sifting procedure in BB84 quantum key distribution protocol. Success is reported when the output polarization state matches the input one. Otherwise, it is considered as an error. To protect against detector attacks in which the terminal's measurement basis can be probed or controlled \cite{Bozzio2021}, the bank randomly assigns an outcome in cases where both detectors click \cite{Beaudry08}. The error rate is calculated as the ratio of errors to the total number of detection events. 

We conducted the experiment for four different mean photon numbers, namely $\mu=0.5$, 1, 1.5, and 2. The generated secret key is composed of 28 random polarization states. One state is encoded every 35 $\mu$s during the fraction of the experimental cycle dedicated to storage. In this implementation the key is not changed from one cycle to the other, and we repeat the sequence 4000 times to acquire sufficient statistics.

Experimental error rates are presented in Fig. \ref{figure4} as a function of the mean photon number. The grey points indicate data collected without storage, similar to an on-the-fly implementation, while the blue points correspond to the complete protocol with intermediate storage in our quantum memory. In the case of storage, the hatched area denotes the insecure regime, and the light red area corresponds to the threshold accounting for the error bar on the average memory efficiency across the four photon numbers $\eta=(77\pm2)\%$.

The implementation was optimized without storage, aiming for polarization purities close to unity and with similar values across the different states. This required a specific strategy for Pockell cell driving, combined with high-quality free-space optics and polarization filtering, along with precise adjustment of the phase difference between the two arms of the interferometer used for dual-rail conversion (see Supplementary Information). The resulting error rates without storage are low, with $\varepsilon=(0.36\pm0.08)\%$ for $\mu=0.5$, $\varepsilon=(0.36\pm0.06)\%$ for $\mu=1$, $\varepsilon=(0.25\pm0.04)\%$ for $\mu=1.5$ and $\varepsilon=(0.29\pm0.04)\%$ for $\mu=2$. These values are in agreement with polarization fidelities of about 99.5\%. The achieved error rates are smaller by more than an order of magnitude compared to previous on-the-fly implementations \cite{Bozzio2018,Guan2018}.

With this, we can now consider the complete implementation including the memory layer, represented by the blue points in Fig. \ref{figure4}. The error rates amount to $\varepsilon=(1.84\pm0.15)\%$ for $\mu=0.5$, $\varepsilon=(0.78\pm0.07)\%$ for $\mu=1$, $\varepsilon=(0.69\pm0.06)\%$ for $\mu=1.5$ and $\varepsilon=(0.87\pm0.05)\%$ for $\mu=2$. As expected, the error rates are higher compared to the on-the-fly case. This is due to an additional constant background noise coming from residual leakage and scattering of the control beam into the detection modes (see Supplementary Information). 

These results demonstrate that our implementation can operate effectively within the secure regime. For instance, the data for $\mu=1$, also depicted in Fig. \ref{figure2}, is below the threshold by about 20 standard deviations. For $\mu=2$, the threshold is more difficult to beat as the acceptable rate significantly decreases with the increased multi-photon components that enable additional attacks.  

The results shown in Fig. \ref{figure4} correspond to a storage time of 1 $\mu$s. Memory efficiency decreases with the storage time due to the residual magnetic fields in our setup (see Supplementary Information). This leads to two consequences: the secure operational range reduces as the threshold decreases with the efficiency, and the error rate increases due to a reduced signal-to-noise ratio. Given our $1/e^2$ memory lifetime of 15 $\mu$s and a constant background level, the maximum storage time for which the realization with $\mu=1$ remains secure is calculated to be 6 $\mu$s, equivalent to a  light propagation distance of 1.2 km in an optical fibre. This value is not a limitation of our cold-atom platform as various additional methods could extend the memory lifetime, up to the subsecond regime \cite{Zhao2009,Yang2016}.

Another important aspect to address is advancing beyond few-mode quantum memories. In the context of the quantum money scheme, multimode memories could enable the simultaneous storage and retrieval of all the qubits. Depending on the physical platform, various multiplexing methods can be used \cite{Heshami2016}. For cold-atom-based devices, the spatial degree of freedom presents a promising avenue for achieving large capacity \cite{Pu2017,Parniak2017,Zhang2024}. Yet, preserving the required high efficiency in these implementations remains a subject of active investigation.

In conclusion, our work provides the first realization of a quantum cryptographic primitive that integrates an intermediate quantum memory layer. The protocol we chose imposes stringent performance requirements on the memory to operate in a secure regime. Using a high-efficiency, low-noise cold-atom-based quantum memory, alongside an optimized photonic setup, we successfully implemented a provably unforgeable quantum money scheme. This result highlights that, beyond entanglement distribution, the availability of such quantum memories unlock new possibilities for implementing protocols that were previously considered out of reach.

We anticipate that our demonstration can be extended to a wide range of quantum protocols, including in two-way quantum communication complexity and fundamental cryptographic primitives requiring storage over communication networks. Potential extensions include prepare-and-measure schemes like coin flipping \cite{Neves2023}, secure multiparty protocols such as secret sharing \cite{Bell2014,Lu2016}, and anonymous transmission \cite{Christandl2005,Unnikrishnan2019}. Beyond cryptographic applications, the successful validation of quantum memory technology under these demanding conditions paves the way for its broader role as a core component in quantum interconnects \cite{PRX}, laying further the groundwork for functional quantum networks.

\vspace{0.5cm}
\noindent \textbf{Funding:} This work was supported by the French National Research Agency via the France 2030 projects QMemo (ANR-22-PETQ-0010) and QCommTestbed (ANR-22-PETQ-0011), and by the European Union's Horizon Europe research and innovation programme via the QIA-Phase 1 project (101102140) and the QSNP project (101114043). H.M. acknowledges support from Region Ile-de-France in the framework of DIM SIRTEQ, T.N. from the EU (Marie Curie fellowship 101029591), and M.B. from the Austrian Science Fund FWF 42 via F7113 (BeyondC) and the AFOSR via FA9550-21-1-0355 (QTRUST). J.L. is a member of the Institut Universitaire de France.

\clearpage
\renewcommand{\thefigure}{S\arabic{figure}}
\renewcommand{\thetable}{S\arabic{table}}
\setcounter{figure}{0}

\onecolumngrid

\centerline{\textbf{SUPPLEMENTARY INFORMATION}}

\section{Encoding of the polarization states}
Random numbers are generated using a quantum random number generator (Quantis-PCI-4, ID Quantique) and converted into voltages by a digital-to-analog converter (USB-6363, National Instruments). These signals are then amplified (PZD350A, Trek) and applied to a Pockels cell (LM0202, LINOS) to create the four polarization states, with voltages ranging from 0 to 450~V. The settling time of the amplifier is about 30 $\mu$s. Importantly, this settling behaviour slightly varies with the specific voltage transitions, as larger overshoots from larger jumps take time to die out, affecting the encoded polarisation as it depends on the voltage jump for which the system is optimized. This limits the fidelity that can be obtained for every state, leading typically to a potential 1\% error rate in the protocol. To mitigate this issue, we introduced a intermediate 10-$\mu$s voltage plateau at 250~V  between steps. This ensures a well-defined and smoother transition between states and drastically enhances the overall fidelity of the encoded polarizations, as required by the stringent secure operation of the protocol. 

\section{High-efficiency cold-atom-based memory}
Cesium atoms are trapped in a compressed quasi two-dimensional magneto-optical trap, with a length of up to 3 cm and an optical depth of up to 400. The mapping process is based on dynamic electromagnetically-induced transparency. The power of the control beam is around 2~mW with a waist of 1~mm. The signal pulse has a gaussian temporal profile with a full width at half maximum of 230~ns. Control and signal beams are almost collinear with an angle of 1$^\circ$ and need to have the same circular polarization when they reach the atoms, requiring specific polarization transformations as shown in Fig. 3 of the main text. During the storage phase, the magnetic field is turned off and residual fields are dynamically cancelled using additional coils. The average broadening measured by microwave spectroscopy is of the order of 50 kHz (full width at half maximum), resulting in a memory lifetime of 15$\, \mu$s. For the detection stage, which is part of the vendor setup, two Fabry-Perot cavities (FPE001A, Quantaser) are used to filter out the control beam leakage, with a typical rejection of 40~dB at 9.2 GHz and a transmission of about 70\% for the signal. This filtering is critical for the experiment as very low error rates are required. Finally, photons are detected with single-photon counting modules (AQRH 14-FC, Excelitas). 

\section{Experimental sequence}

\begin{figure}[b!]
\label{fig1_SM}
\begin{minipage}[b]{0.3\linewidth}
\centering
\caption{\textbf{Timing diagram of the experiment.} A very elongated MOT is first loaded, followed by a compression and a polarization gradient cooling (PGC) stage. Then, during a 2-ms phase, the atomic ensemble is used for the cryptographic protocol implementation, with successive storage and retrieval of the optical qubits. }
\vspace{1cm}
\end{minipage}
\hspace{0.1cm}
\begin{minipage}[b]{0.67\linewidth}
\centering
\includegraphics[width=0.98\columnwidth]{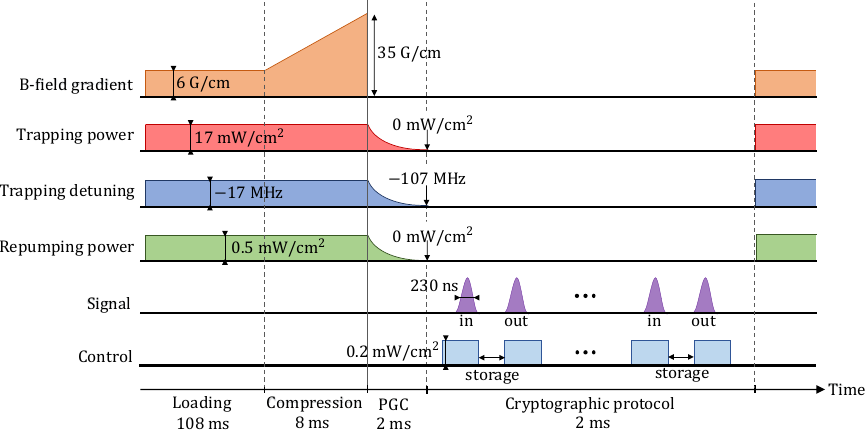}
\end{minipage}
\end{figure}

The experiment is running on a cycle of 120~ms, as described in Fig. S1. First, a loading phase of 108 ms is performed, in which all the parameters are set in constant mode. The elongated magneto-optical trap (MOT) is based on two pairs of rectangular coils, resulting in a magnetic field gradient in the transverse axis of 6~G/cm and longitudinal one of about 0.4 G/cm. The total trapping power is 350 mW with an intensity of 17 mW/cm$^2$. The trapping beam is red detuned by 17 MHz from the $\big|6S_{1/2},F=4\bigr \rangle$ $\rightarrow$ $\big|6S_{1/2},F'=5\bigr \rangle$ cycling transition. The total repumping power is 4~mW with an intensity of 0.2 mW/cm$^2$. After MOT loading, a compression phase is initiated and the magnetic field gradient is increased linearly from 6 to 35~G/cm in 8 ms. At the end of the compression stage, the magnetic field is switched off and we perform polarization gradient cooling (PGC) during 2 ms. To do so, we ramp down the trapping and repumping power to zero, while trapping detuning is increased from -17 to -107 MHz. This phase results in a final temperature for the atoms of about 20 $\mu$K, determined by a time-of-flight measurement. The extinction time for the trapping power is 1 ms longer than the one for the repumping, preparing all the atoms in the $\big|6S_{1/2},F=3\bigr \rangle$ ground state. The achieved OD is about 400. Three pairs of bias coils are used to cancel the residual magnetic field.

After the sequence dedicated to loading and cooling, the cryptographic protocol is performed. The random polarizations are encoded on the signal pulse with a Pockels cell. They are stored and retrieved in and out of the memory using dynamical EIT. The control beam is turned on before the arrival of the signal pulse on the atoms and turned off when the pulse is entirely compressed into the cloud. After a defined storage time, the control beam is turned on again to retrieve the initial signal pulse. For this protocol, the storage time is about 1 $\mu$s. The FWHM of the signal pulse was set to 230~ns and the control intensity to 0.2 mW/cm$^2$. The storage-and-retrieval process is repeated 28 times during a MOT cycle.  

\section{Security threshold calculation}
A dishonest client will attempt to double-spend the quantum money in their possession. For this attack to succeed with two distinct verifiers, their strategy will involve some form of cloning operation applied to the states sent by the bank. While the random encoding of these states  in conjugate bases inherently forbids perfect quantum cloning, experimental imperfections such as finite efficiencies, multiphoton contributions and depolarizing channels can all be exploited by the dishonest client to increase their success probability. Our security proof searches for the optimal cheating strategy allowed by the quantum mechanics, accounting for the experimental noise and loss. We model the client's strategy as a completely-positive trace-preserving quantum map and apply linear constraints arising from the honest protocol. Using semidefinite programming, we derive upper bounds on the loss and noise allowed in the experiment. If the imperfections exceed these upper bounds, a dishonest client can perfectly cheat, making the protocol insecure.

\section{Detailed security analysis}
 
Here, we provide the necessary tools required to understand the practical security analysis of our quantum money demonstration. We start by deriving the expressions for the weak coherent states used in our experiment, followed by some mathematical preliminaries on semidefinite programming (SDP) and Choi's theorem on completely positive maps. Finally, we detail the derivation of our practical security thresholds. 

\subsubsection{Modelling of the weak coherent states}
Coherent states may be expressed as a Poisson-distributed superposition of photon number states:

\begin{equation}
    \ket{\alpha} = \sum_{n=0}^{\infty} e^{-\frac{|\alpha|^2}{2}}\frac{\alpha^n}{\sqrt{n!}}\ket{n} =\sum_{n=0}^{\infty} C_\alpha\left(n\right)\ket{n},
\end{equation}
where $\{\ket{n}\}$ denote the photon number states and $\alpha$ is the coherent state amplitude. Although our experiment is performed with polarization qubits from the set $\{\ket{H},\ket{\sigma^+},\ket{V},\ket{\sigma^-}\}$, we perform our security analysis with the equivalent set $\{\ket{D},\ket{\sigma^+},\ket{A},\ket{\sigma^-}\}$, which elegantly maps onto two-mode weak coherent states as:

\begin{equation}
\ket{\alpha_k} = \Ket{e^{i\theta}\frac{\alpha}{\sqrt{2}}}\otimes\Ket{e^{i(\theta+\phi_k)}\frac{\alpha}{\sqrt{2}}},
\label{eq:globalphase}
\end{equation}
where $\theta\in[0,2\pi]$ is a global phase and $\phi_k\in\{0,\pi/2,2\pi,3\pi/2\}$ is the relative phase between the two modes, which can take one of four values depending on $k\in\{0,1,2,3\}$. 

In a dishonest setting, an adversary must access $\phi_k$ to unveil the information encoded in the states. In order to decrease the impact of discrimination attacks exploiting a global phase reference, we assume that the phase $\theta$ from Eq. (\ref{eq:globalphase}) is uniformly randomized over $[0,2\pi]$. In practice, phase randomization can be achieved using for instance laser gain switching \cite{Kobayashi2014} or active phase modulation with a sufficient number of discrete phases \cite{Cao2015}.

Under this assumption, integrating $\ket{e^{i\theta}\alpha}$ over all possible values of $\theta$ reduces the forwarded state to a classical mixture of number states \cite{Lo2005}:
\begin{equation}
    \frac{1}{2\pi}\int_{0}^{2\pi} \ket{\sqrt{\mu} e^{i\theta}}\bra{\sqrt{\mu} e^{i\theta}}d\theta = e^{-\mu}\sum_{n=0}^{\infty}\frac{\mu^n}{n!}\ket{n}\bra{n},
\end{equation}
where $\mu=|\alpha|^2$ is the average photon number per state. As coherence between number states vanishes, our security proof may simply proceed according to the result of quantum non-demolition (QND) photon number measurements. When the state contains $0$ photons, no information can be accessed by the adversary. When it contains $1$ photon, the qubit security proof may be applied. When it contains $2$ or more photons, perfect cheating is assumed.

This decomposition allows to express the phase-randomized states $\{\rho_k\}$ in a $7$-dimensional orthonormal basis $\{\ket{v},\ket{H},\ket{V},\ket{m_0},\ket{m_1},\ket{m_2},\ket{m_3}\}$, where $\ket{v}$ is the vacuum state, $\ket{H}$ and $\ket{V}$ span a polarization qubit space, and $\ket{m_i}$ constitute the four orthonormal outcomes which materialize the four perfectly distinguishable states in the multiphoton subspace. Our four phase-randomized coherent states may then be written as the following density matrices \cite{Bozzio2019}:
\begin{equation}
\begin{aligned}
    \rho_0 &= P_\mu(0)\:\ketbra{v}{v}+P_\mu(1)\:\ketbra{H}{H}+P_\mu(\geqslant 2)\:\ketbra{m_0}{m_0}\\
    \rho_1 &= P_\mu(0)\:\ketbra{v}{v}+P_\mu(1)\:\ketbra{\sigma^+}{\sigma^+}+P_\mu(\geqslant 2)\:\ketbra{m_1}{m_1}\\
    \rho_2 &= P_\mu(0)\:\ketbra{v}{v}+P_\mu(1)\:\ketbra{V}{V}+P_\mu(\geqslant 2)\:\ketbra{m_2}{m_2}\\
    \rho_3 &= P_\mu(0)\:\ketbra{v}{v}+P_\mu(1)\:\ketbra{\sigma^-}{\sigma^-}+P_\mu(\geqslant 2)\:\ketbra{m_3}{m_3},
\end{aligned}\label{eq:randomphase}
\end{equation}
where $\{\ket{D},\ket{\sigma^+},\ket{A},\ket{\sigma^-}\}$ denote the usual superpositions in the space spanned by $\{\ket{H},\ket{V}\}$ and the Poisson coefficients are given by:
\begin{equation}
\begin{aligned}
    P_\mu(0)&= e^{-\mu},&
    P_\mu(1) &= \mu e^{-\mu},&
    P_\mu(\geqslant 2) &= 1-(1+\mu)e^{-\mu}.
\end{aligned}\label{eq:PDScoeffs}
\end{equation}

\subsubsection{Semidefinite programming}\label{sec:sdp}

 Quantum-cryptographic security proofs optimize over semidefinite positive objects to derive bounds on an adversary's cheating probability. These objects can be density matrices, measurement operators, or more general completely positive trace-preserving (CPTP) maps. Semidefinite programming provides a suitable framework for this, as it allows to optimize over semidefinite positive variables, given linear constraints \cite{Watrous2011,Molina2013}.

A semidefinite program may be defined as a triple $\left(\Lambda,F,C\right)$ where $\Lambda$ is a Hermitian-preserving CPTP map, and $F$ and $C$ are Hermitian operators living in complex Hilbert spaces $\mathcal{H}_F$ and $\mathcal{H}_C$, respectively. The primal problem maximizes a \textit{primal objective function}, $\Tr\left(F^{\dagger}X\right)$,  over all positive semidefinite variables $X$, given a set of linear constraints expressed as a function of $C$:

\begin{equation}
\begin{aligned}
\text{maximize} &&& \Tr\left(F^{\dagger}X\right)\\
\text{s.t.}  &&& \Lambda(X) = C\\
&&& X \geqslant 0.
\end{aligned}\label{primal}
\end{equation}
If it exists, the operator $X$ which maximizes $\Tr\left(F^{\dagger}X\right)$ given these constraints is the \textit{primal optimal solution}, and the corresponding value of $\Tr\left(F^{\dagger}X\right)$ is the  \textit{primal optimal value}.

Semidefinite programs present an elegant dual structure, which associates a dual minimization problem to each primal maximization problem. Effectively, the new dual variable(s) $Y$ may be understood as the Lagrange multipliers associated with the constraints of the primal problem (one for each constraint). The dual problem associated with \eqref{primal} may then be written as:

\begin{equation}
\begin{aligned}
\text{minimize} &&& \Tr\left(C^{\dagger}Y\right)\\
\text{s.t.}  &&& \Lambda^*(Y) - F \geqslant 0\\
&&& Y = Y^{\dagger}.
\end{aligned}\label{duall}
\end{equation}
 Similarly to the primal problem, the operator $Y$ which minimizes $\Tr\left(C^{\dagger}Y\right)$ given these constraints, if it exists, is the \textit{dual optimal solution}, and the corresponding value of $\Tr\left(C^{\dagger}Y\right)$ is the  \textit{dual optimal value}.

The Lagrange multiplier method allows to find the local extremum of a constrained function. The optimal value $s_p$ of the primal problem therefore lower bounds the optimal value $s_d$ of the dual problem, while the optimal value of the dual upper bounds that of the primal. This property is known as \textit{weak duality}, and may be simply expressed as:

\begin{equation}
    s_p\leqslant s_d.
\end{equation}
In many quantum-cryptographic applications, we wish to ensure that the upper bound derived in the primal problem is \textit{tight}, i.e. that the local maximum is in fact a global maximum for the objective function. The dual problem will help to prove this when there exists \textit{strong duality}:

\begin{equation}
    s_p= s_d.
\end{equation}

\subsubsection{Choi's theorem on completely positive maps}\label{sec:choi}

We now recall Choi's theorem on completely positive maps, which establishes useful equivalences between properties of linear maps and those of density operators. Let us consider a tensor product of two $d$-dimensional Hilbert spaces $\mathcal{H}=\mathcal{H}_1^d\otimes\mathcal{H}_2^d$, and then define the maximally entangled state $\ket{\Phi^{+}}\bra{\Phi^{+}}$ on $\mathcal{H}$ as
\begin{equation}
\ket{\Phi^{+}}\bra{\Phi^{+}} = \frac{1}{d}\sum_{i,j=1}^{d} \ket{i}\bra{j}\otimes\ket{i}\bra{j}
\end{equation}
We introduce a completely positive linear map $\Lambda: \mathcal{H}_1^d \rightarrow \mathcal{H}_3^{d'}$, and define the Choi-Jamiolkowski operator $J(\Lambda) : \mathcal{H}_1^d\otimes\mathcal{H}_2^d \rightarrow \mathcal{H}_3^{d'}\otimes\mathcal{H}_2^d$ as the operator which applies $\Lambda$ to the first half of the maximally entangled state $\ket{\Phi^{+}}\bra{\Phi^{+}}$:
\begin{equation}
J(\Lambda) = \frac{1}{d}\sum_{i,j=1}^{d} \Lambda(\ket{i}\bra{j})\otimes\ket{i}\bra{j}.
\label{eq:choi}
\end{equation}
Choi's theorem then states that $\Lambda$ is completely positive if and only if
$J(\Lambda)$ is positive semidefinite. We also have that $\Lambda$ is a trace-preserving map if and only if $\Tr_{\mathcal{H}_3^{d'}}(J(\Lambda)) = \mathbbb{1}_{\mathcal{H}_2^d}$ \cite{Watrous2011,Molina2013}. These properties are implemented as constraints in our optimization problem.

\subsubsection{Threshold calculation}

The calculations closely follow those from \cite{Bozzio2019}, considering a quantum money scheme with quantum verification. In such a scheme, a successful forging attack is one in which two copies of the quantum money state are accepted at two spatially separated verification points. 
 
 Let $\Lambda$ be the optimal adversarial map which produces two copies (living in $\mathcal{H}_1\otimes\mathcal{H}_2$) of the original quantum money state living in $\mathcal{H}_{\text{ini}}$:

\begin{equation}
\rho_{\text{ini}} = \frac{1}{4}\sum_{k=0}^3 \rho_k.
\end{equation}

By imposing a condition on the terminal's postprocessing, consisting of assigning a random measurement outcome $\ket{0}$ or $\ket{1}$ to any double click and declaring a flag $\ket{\varnothing}$ when no detection is registered \cite{Beaudry2008}, one can express the threshold detector measurement operators in a 3-dimensional Hilbert space spanned by $\{\ket{0},\ket{1},\ket{\varnothing}\}$. The probability that a verifier declares an incorrect measurement on the first copy is given by:

\begin{equation}
V_0 = \Tr\sum_{k=0}^3\left(\frac{1}{2}\ket{\beta_{k}^\perp}\bra{\beta_{k}^\perp}\otimes \mathbbb{1}\right) \Lambda\left(\frac{1}{4}\rho_k\right),
\end{equation} 
while for the second copy this reads:

\begin{equation}
V_1=\Tr\sum_{k=0}^3\left(\mathbbb{1}\otimes\frac{1}{2}\ket{\beta_{k}^\perp}\bra{\beta_{k}^\perp}\right) \Lambda\left(\frac{1}{4}\rho_k\right),
\end{equation}
where $\ket{\beta_{k}}$ is the squashed qubit associated with the original state $\rho_k$, i.e.\@ $\ket{\beta_0} = \ket{H}$, $\ket{\beta_1} = \ket{\sigma^+}$, $\ket{\beta_2} = \ket{V}$, $\ket{\beta_3} = \ket{\sigma^-}$, and $\ket{\beta_{k}^\perp}$ is its orthogonal qubit state. The factor $1/4$ indicates that each $\rho_k$ is equally likely to occur, while $1/2$ accounts for the verifier's random measurement basis choice. Using the Choi formalism from Section \ref{sec:choi}, we may rewrite these expressions as $V_0 =\Tr\left(E_0(\mu)J(\Lambda)\right)$ and $V_1 = \Tr\left(E_1(\mu)J(\Lambda)\right)$, where $E_0(\mu)$ and $E_1(\mu)$ are the \textit{error operators}:

\begin{equation}
\begin{aligned}
   E_0(\mu)=&\frac{1}{4}\sum_{k=0}^{3}\frac{1}{2} \ket{\beta_{k}^\perp}\bra{\beta_{k}^\perp}\otimes\mathbbb{1}\otimes\overline{\rho_k}, \\
   E_1(\mu)=&\frac{1}{4}\sum_{k=0}^{3}\mathbbb{1}\otimes\frac{1}{2} \ket{\beta_{k}^\perp}\bra{\beta_{k}^\perp}\otimes \overline{\rho_k} .
\end{aligned}
\end{equation}

Following a similar method, the probability that the first (resp. the second) verifier registers a no-detection event for the first (resp. second) copy reads $\Tr\left(L_0(\mu)J(\Lambda)\right)$ (resp. $\Tr\left(L_1(\mu)J(\Lambda)\right)$), where $L_0(\mu)$ and $L_1(\mu)$ are the \textit{loss operators}, which contain the projection onto the state $\ket{\varnothing}$:
\begin{equation}
\begin{aligned}
L_0(\mu) = \frac{1}{4}\sum_{k=0}^3 \ket{\varnothing}\bra{\varnothing}\otimes\mathbbb{1} \otimes\overline{\rho_k}, \\
L_1(\mu) = \frac{1}{4}\sum_{k=0}^3
\mathbbb{1}\otimes\ket{\varnothing}\bra{\varnothing} \otimes\overline{\rho_k} .
\label{will}
\end{aligned}
\end{equation}
We now search for the optimal cloning map $\Lambda$ that minimizes the noise that the adversary must introduce for both copies given a fixed combined channel and detection loss $ e^{-\eta_m\mu}$, where $\eta_m$ is the combined storage/retrieval efficiency of our quantum memory. We cast this problem in the following SDP for an attack on a single quantum state, and solve it using the SDPT3 solver of the MATLAB CVX package: 
\begin{equation}
\begin{aligned}
\min\quad& \Tr\left(E_0(\mu) J(\Lambda) \right) \\
\text{s.t. }&  \Tr_{\mathcal{H}_1\otimes\mathcal{H}_2}\left(J(\Lambda)\right) = \mathbbb{1}_{\mathcal{H}_{\text{ini}}} \\
    &\Tr\left(E_0(\mu) J(\Lambda) \right) \geqslant \Tr\left(E_1(\mu) J(\Lambda) \right) \\
    &\Tr\left(L_0(\mu) J(\Lambda) \right) \leqslant e^{-\eta_m\mu} \\
    &\Tr\left(L_1(\mu) J(\Lambda) \right) \leqslant e^{-\eta_m\mu} \\
    &J(\Lambda) \geqslant 0
\end{aligned}\label{eq:noisetol}
\end{equation}
The first constraint imposes that $\Lambda$ is trace-preserving, the second imposes that the error rate measured for the first copy is at least equal to the one measured for the second copy, the third and fourth impose that the losses measured for tokens $1$ and $2$ do not exceed the expected honest losses, and the fifth imposes that $\Lambda$ is completely positive. Note that the optimal value obtained from Problem (\ref{eq:noisetol}) should be divided by the probability of detecting at least one photon, given by $\left(1-e^{-\eta_m\mu}\right)$. 

Since strong duality holds for our problem \cite{Bozzio2019}, this lower bound is in fact optimal. Furthermore, following the product rule of semidefinite programs and the arguments from \cite{Bozzio2019}, the adversary cannot succeed better by performing a general attack on the full tensor product of the $N$ states contained in the quantum money state.

\end{document}